\documentstyle[12pt]{article}
\begin{document}
\title{Gravitational Bending of Light with Frequency Shifts}
\author{P.D. Morley \\XonTech, Inc. \\6862 Hayvenhurst Ave.
\\ Van Nuys, CA 91406}
\maketitle
\begin{abstract}
Non-static gravitational fields generally introduce frequency shifts when
bending light. In this paper, I discuss the frequency shifts induced in the
bending of light by moving masses. As examples, I treat the recently
discovered high-velocity pulsar PSR 2224+65 and a typical Einstein ring.
\end{abstract}

\vspace{.25in}

\begin{center} PACS Numbers: 97.60Gb, 97.60Jd \end{center}

\vspace{.25in}
The four Jupiter probes Pioneers 10, 11 and Voyagers 1,2 have all reached
solar system escape velocities by the employment of the "sling-shot"
effect. This effect involved sending the spacecraft in a trajectory such
that it received a gravitational boost from Jupiter's motion. One can
say that during the time of the satellite's proximity to Jupiter, Jupiter
carried the craft along its orbit. An equivalent view is to treat the
satellite-Jupiter encounter as a two-body scattering event with the
spacecraft leaving with additional kinetic energy at Jupiter's expense.
The analogous situation will occur for light: light bending by a moving
object will introduce a frequency shift. In effect, the light is being carried
along by the moving mass.

One could discuss the light effect by employing a complicated
time-dependent gravitational potential. Equivalently, the same answer is
obtained by the use of the scattering formalism. Since the latter is more
transparent, we will use that method.

First, it is instructive to see how the spacecraft slingshot method
works. In the center of momentum frame (CM) the spacecraft with mass,
$m_{s}$, goes down and up the potential well with no net change of
kinetic energy. The CM moves with velocity $\vec{v}_{p}$ with respect
to the heliocentric frame of reference (which we will call the lab frame LF),
with $\vec{v}_{p}$ the velocity of the planet. In the LF, Galilean
invariance requires that the incoming velocity will be $\vec{v}_{i}
= \vec{v}_{i}^{/} + \vec{v}_{p}$, with $\vec{v}_{i}^{/}$ being the
incoming CM velocity, while the outgoing velocity will be $\vec{v}_{f}
= \vec{v}_{f}^{/} + \vec{v}_{p}$. The change in kinetic energy $\Delta E$
in the LF is
\begin{equation} \Delta E = m_{s}(\vec{v}_{f}^{/} - \vec{v}_{i}^{/})
\cdot \vec{v}_{p} = m_{s} \Delta \vec{v}^{/} \cdot \vec{v}_{p} =
m_{s} \Delta \vec{v} \cdot \vec{v}_{p}. \end{equation}
In this equation, $\Delta \vec{v}^{/} = \Delta \vec{v}$ is the change
of velocity of the spacecraft due to the slingshot motion and is the
same quantity in both frames. Thus, the craft can either gain or lose
energy depending on the trajectory. The maximum possible energy gain
occurs when the spacecraft scatters head on with the planet in the CM
and the slingshot trajectory carries it outgoing in the opposite direction.

To compute the frequency shift for light, we consider the following steps:

[A]. In the lab frame LF, the incoming photon's energy is $\nu$, and it
makes an angle $\theta_{in}$ with $\vec{v}$, where $\vec{v}$ is the lab
velocity of the bending mass.

[B]. In the CM, this incoming photon is Doppler shifted to a frequency
$\nu^{/}$, where the relations between these CM variables and the LF
variables are given by the usual Doppler formulae $(\beta = v/c)$:
\begin{equation} \nu = \frac{\nu^{/} \sqrt{1-\beta^{2}}}{1- \beta \cos
\theta_{in}}, \; \cos \theta_{in}^{/} = \frac{\cos \theta_{in} - \beta}
{1 - \beta \cos \theta_{in}}. \end{equation}

[C]. The bending of light in the CM changes $\theta_{in}^{/}$ to
$\theta_{out}^{/}$ but does not modify $\nu^{/}$. General relativity [1]
provides the formula for the change in angle:
\begin{equation} \delta \theta^{/} = | \theta_{in}^{/} - \theta_{out}^{/} |
= \frac{4MG}{\rho c^{2}} , \end{equation}
where $\rho$ is the distance of closest approach, and $M$ is the bending
mass. After bending, the new CM angle is $\theta_{out}^{/}$.

[D]. The outgoing photon frequency is Doppler shifted to $\nu^{//}$ in the
LF, with angle $\theta_{out}$:
\begin{equation} \cos \theta_{out} = \frac{\cos \theta_{out}^{/} + \beta}
{1 + \beta \cos \theta_{out}^{/}}, \; \nu^{//} = \frac{\nu^{/}
\sqrt{1 - \beta^{2}}}{1- \beta \cos \theta_{out}} = \frac{\nu (1 - \beta
\cos \theta_{in} )}{1 - \beta \cos \theta_{out}} .  \end{equation}
The fractional frequency shift in the LF is
\begin{equation} \frac{\Delta \nu}{\nu} = \frac{\nu^{//} - \nu}
{\nu} = \kappa-1, \; \kappa = \frac{1- \beta \cos \theta_{in}}
{1-\beta \cos \theta_{out}}. \end{equation}
These results give the answer to the frequency shift of light induced
when a light beam undergoes gravitational bending. The angles
$\theta_{in}, \theta_{out}$ are the incoming and outgoing photon angles
made with respect to $\vec{v}$ in the lab frame LF.

Consider the specific examples of the high velocity pulsar PSR 2224+65
and a typical Einstein ring.

The pulsar has a measured [2] transverse speed $\geq$ 800 km/s due to
either an asymmetric supernova explosion or a stellar three-body
scattering event. I assume it to be a neutron star of mass $M \sim 1.4
M_{\odot}$ and radius $R \sim $ 10 km. For the purpose of computing
the frequency shifts, we take the pulsar's velocity vector to be
entirely transverse, so that $\theta_{out} = \pi /2$.

Given $\rho$ (which of course must be greater than the radius of the
pulsar), there are two different values of $\theta_{in}$ in (3): one
is associated with a red-shift trajectory and a fractional
frequency shift $\Delta \nu_{r}/ \nu$, and the other is associated
with a blue-shift trajectory and a fractional frequency shift
$\Delta \nu_{b}/ \nu$. These quantities are given in Table 1. In Fig. 1,
the geometrical arrangements are given, showing the (r) and (b)
trajectories.

To see how the table was compiled, I briefly go over the first line,
with $\rho =$ 10 km, so that light just grazes the pulsar. Since
$\theta_{out} = \pi /2$, we have from (4) that $\cos \theta_{out}^{/}
= -\beta = -2.67 \times 10^{-3}$. The change in angle is given by (3):
$\delta \theta^{/}=0.8269$; the two incoming angles in the CM are
$\theta_{r}^{/}=0.7466$ and $\theta_{b}^{/}=2.4004$. Then (2) gives
$\cos \theta_{r}=0.7353$, $\cos \theta_{b} =-0.7364$, so the
two trajectories have LF incoming angles $\theta_{r,in}=0.7447$,
$\theta_{b,in}=2.3986$. These have the computed fractional frequency shifts
given in Table 1; the rest of the table is computed in the same way.

Next, as the second example, I treat the case of a special gravitational
lens, the Einstein ring. This ring results when the source, lens and
observer are all on one line. There are presently three known cases of
such imaging: MG 1131+0456 [3], MG 1654+1346 [4] and PKS 1830-211 [5]. The
geometry of the ring is illustrated in Fig.2. For simplicity, we have put
the velocity of the lens $\vec{v}$ in the x-z plane with a polar angle
$\theta = \pi /3$. It is obvious that the azimuthal angle of the light
ray changes from 0 to 2$\pi$; the angle formed with $\vec{v}$ also
changes, and so the frequency shift depends on the azimuthal angle
$\psi$. We choose typical values $\psi \sim 10^{-3}$ rad. The velocity
dispersion for galaxies is $\sim 250$ km/s, and we adopt this value for
the lens speed.

Table 2 gives some results. The frequency shift changes sign depending
on azimuthal angle and has a value of about one part in a million.
That is, a 15GHz radio line will experience a 15kHz shift due to the
movement of the gravitational lens.

It is clear that any object which is capable of bending light can
introduce a frequency shift due to its motion, including a black hole.
A blue shift indicates that the light ray has extracted energy from the
translational energy of the bending object while a red shift indicates
that the light ray has added to this energy.

The author was a visitor to the Center for Particle Physics at U.T.
Austin and would like to thank its Director for the hospitality extended
to him.

\newpage
\begin{table}
\begin{tabular}{ccc}
\hspace{1.0in} $\rho$ & $\Delta\nu_{r}/ \nu$ & $\Delta \nu_{b} / \nu$   \\
\mbox{} & \mbox{} & \mbox{} \\
\hspace{1.0in} 10 km & -1.962$\times 10^{-3}$ & +1.965$\times 10^{-3}$  \\
\hspace{1.0in} 50 km & -4.392$\times 10^{-4}$ & +4.394$\times 10^{-4}$  \\
\hspace{1.0in} 100 km & -2.204$\times 10^{-4}$ & +2.204$\times 10^{-4}$ \\
\hspace{1.0in} 1.0 R$_{\odot}$ & -3.17$\times 10^{-8}$ & +3.17$\times 10^{-8}$
\\
\hspace{1.0in} 0.1 A.\ U. & -1.4$\times 10^{-9}$ & +1.4$\times 10^{-9} $

\end{tabular}

\caption{Micro-lensing off PSR 2224+65.}
\end{table}

\vspace{2.0in}
\mbox{}

\begin{table}
\begin{tabular}{cc}
\hspace{1.0in} Azimuthal Angle & $\Delta \nu / \nu $ \\
\mbox{} & \mbox{} \\
\hspace{1.0in} 0 & $-1.44 \times 10^{-6} $ \\
\hspace{1.0in} $\pi / 4 $ & $-1.02 \times 10^{-6} $ \\
\hspace{1.0in} $\pi /2 $ & no shift \\
\hspace{1.0in} $3 \pi /4 $ & $+1.02 \times 10^{-6}$ \\
\hspace{1.0in} $\pi$  & $1.44 \times 10^{-6}$ \\
\hspace{1.0in} $5 \pi /4$ & $+1.02\times10^{-6}$ \\
\hspace{1.0in} $3\pi /2$ & no shift \\
\hspace{1.0in} $7\pi /4$ & $-1.02\times10^{-6} $

\end{tabular}

\caption{Fractional frequency shifts of an Einstein ring verses azimuthal.}
\end{table}
\mbox{}
\newpage
\mbox{}
\vspace{1.5in}

\large

\begin{center}  {\bf Figure Captions} \end{center}

\normalsize

\vspace{1.0in}

Fig.\ 1. Micro-lensing of PSR 2224+65. Trajectory (r) is red-shifted
while (b) is blue-shifted. \\

\vspace{.5in}

Fig.\ 2. The Einstein ring has source (s), lens (l) and
observer (o) in a line. For our example, the velocity of the lens $\vec{v}$
lies in the x-z plane.


\newpage
\begin{thebibliography}{99}
\bibitem{} L.\ D. Landau and E.\ M. Lifshitz, {\em The Classical
Theory of Fields} (Pergamon Press, Second Edition, Oxford, 1962).
\bibitem{} J.\ M. Cordes, R.\ W. Romani and S.\ C. Lundgren, Nature
{\bf 362} (1993) 133.
\bibitem{} J.\ N. Hewitt et al., Nature {\bf 333} (1988) 537.
\bibitem{} G.\ I. Langston et al., Astrophys. J. {\bf 97} (1989) 1283.
\bibitem{} D.\ L. Jauncey et al., Nature {\bf 352} (1991) 132.
\end{thebibliography}
\end{document}